\documentclass[twocolumn,pra,showpacs,superscriptaddress,preprintnumbers]{revtex4}

\usepackage{amssymb}
\usepackage{amsmath}
\usepackage{graphicx}
\usepackage{bm}
\usepackage{color}
\usepackage{appendix}
\pdfoutput=1
\begin{document}

\title{Observation of multiramp fractional vortex beams and their properties
in free space}
\author{Jisen Wen}
\affiliation{Department of Physics, Zhejiang University, Hangzhou 310027, China}
\author{Binjie Gao}
\affiliation{Department of Physics, Zhejiang University, Hangzhou 310027, China}
\author{Guiyuan Zhu}
\affiliation{Department of Physics, Zhejiang University, Hangzhou 310027, China}
\author{Yibing Cheng}
\affiliation{Department of Physics, Zhejiang University, Hangzhou 310027, China}
\author{Shi-Yao Zhu}
\affiliation{Department of Physics, Zhejiang University, Hangzhou 310027, China}
\affiliation{Zhejiang Province Key Laboratory of Quantum Technology and Device, Zhejiang
University, Hangzhou 310027, China}
\author{Li-Gang Wang}
\email{sxwlg@yahoo.com}
\affiliation{Department of Physics, Zhejiang University, Hangzhou 310027, China}
\affiliation{Zhejiang Province Key Laboratory of Quantum Technology and Device, Zhejiang
University, Hangzhou 310027, China}

\begin{abstract}
We have experimentally investigated the evolution properties of multiramp
fractional vortex beams (MFVBs) in free space, by using a fundamental
Gaussian beam reflecting from a phase-modulated spatial light modulator. The
issue about the total vortex strength of such MFVBs is systematically
addressed, and our result reveals the dependence of the total vortex
strength depends on both the non-integer topological charge $\alpha $ and
the multiramp number $m$ contained in initial multiramp phase structures. In
the near-field region, vortices contained in MFVBs are unstable and it is
hard to effectively confirm the vortex strength for such fields. However, in
the far-field region, the evolution of vortices in fields becomes stable and
the behavior of vortex strength is confirmed experimentally via measuring
vortex structures by interference method. These findings give us an understanding
of such complex MFVBs and may lead to potential applications in light signal
process and propagation.
\end{abstract}

\date{\today}
\pacs{42.25.-p, 42.25.Bs, 42.25.Hz}
\maketitle

\section{Introduction}
Vortex beams with phase singularities have attracted much attention due to
their potential applications, including optical tweezers, optical
communication, and quantum information. These vortex beams have the phase
structure with the form of $\exp (i\alpha \phi )$, where $\phi $ is the
azimuthal angle and $\alpha $ is a number known as topological charge.
Vortex beams usually have well-defined orbital angular momentum (OAM) when $%
\alpha $ is an integer, and they are also called as integer vortex beams
\cite{Allen1992}. However, when $\alpha $ is a non-integer, then such light
fields are known as fractional vortex beams (FVBs). For an FVB, its initial
phase structure contains only a single mixed screw-edge dislocation \cite%
{Basistiy1995}. Some unique properties and applications of such FVBs have
been proposed, for instance, topologically structured darkness \cite%
{Alperin2007}, fractional OAM momentum entanglement of two photons \cite%
{Oemrawsingh2004,Oemrawsingh2005}, quantum
entanglement between fractional and integer OAM  by single metadevice
\cite{Kun2018}, and quantum digital spiral imaging \cite{Chen2014}.
Due to the fractional OAM carried by FVBs,
the methods of detecting their OAM have also been studied extensively
by using a cylindrical lens \cite{Alperin2016}, dynamic angular double
slits \cite{Zhu2016}, and two-dimensional multifocal array \cite{Deng2019}.
Furthermore, Tkachenko et. al. have tried to generate a ``perfect'' FVBs
\cite{Tkachenko2017}, and the concept of such FVBs is also generalized into
the cases of electron beams \cite{Bandyopadhyay2017} and acoustic waves \cite%
{Jia2018,Hong2015}.

More fundamentally, much attention has been paid to vortex structures of
FVBs, since the distribution of vortices can describe their features. The
unstable evolution of phase structure for FVBs is revealed during their
propagations \cite{Basistiy2004}, thus the total vortex strength is often
adopted to describe their main characteristics \cite{Berry2004,Jesus2012}.
Total vortex strength for FVBs is first presented by Berry \cite{Berry2004},
by assuming an ideal plane wave incident on a non-integer vortex phase
plate. Under this condition, total vortex strength increases by unit only
when topological charge $\alpha $ is slightly lager than any half integer
\cite{Berry2004}. This result is confirmed in some experiments \cite%
{Leach2004,Lee2004,Fang2017}. However, in all practical experiments, light
sources are finite beams not plane-wave fields. There is a different
mechanism on the birth of vortices for FVBs at Fraunhofer zone, which shows
experimentally that total vortex strength for FVBs increases by unit only at
a number slightly larger than an integer \cite{Jesus2012}. Most recently, we
reconsidered this issue on the change of total vortex strength for FVBs at
Fraunhofer zone, and found both theoretically and experimentally that total
vortex strength for FVBs occurs two jumps only when non-integer topological
charge is before and after (but very close to) any even integer number \cite%
{WJS2019}. The incident light sources used in Refs. \cite{Jesus2012,WJS2019}
are finite-width Gaussian beams, which are actually similar to the cases in
other works \cite{Basistiy2004,Leach2004,Lee2004,Fang2017}. In
fact, it is a practical way to utilize a Gaussian beam instead of an ideal plane wave
to be incident on the designed phase plate with screw-edge dislocation or
hologram gratings to generate FVBs. Therefore there are some interesting
questions, how the total vortex strength of a practical FVB changes in free
space and how it changes as $\alpha $ varies. The answer of these questions
are important to the potential application of practical FVBs.

On the other hand, FVBs has only a single mixed screw-edge dislocation, and
they are generalized into a kind of beams with multiple screw-edge
dislocations. For such light fields, one uses $\alpha $ to denote
non-integer topological charge and $m$ to denote a multiramp number
contained in initial multiramp phase structures, thus they are so-called
multiramp fractional vortex beams (MFVBs). Such MFVBs are used for
demonstrating the Hilbert's hotel paradox \cite{Gbur2016}. As stated before,
for a practical MFVB when a finite beam is used as an incident field, its
total vortex strength remains unsolved. An unified explanation is needed for
the vortex strength for both practical FVBs and MFVBs in free space. In
order to answer the above problems, we have both theoretically and
experimentally investigated the propagation properties of MFVBs (including
FVBs) in free space, from their total vortex strength and vortex phase
structures.
\section{Theory}
We start from the propagation properties of MFVBs in free space. In
practical cases, light sources are always finite in transverse dimensions.
Here we consider a fundamental Gaussian beam as an incident beam, and let it
pass through (or reflect from) a phase-modulated spatial light modulator
(SLM). Then the light field at initial plane (i.e., at the plane of SLM) can
be simply given as
\begin{equation}
E_{\text{in}}(r,\phi ,0)=t(\phi )\exp (-\frac{r^{2}}{w_{0}^{2}}),
\label{EINPUT}
\end{equation}%
where $r$ is polar radius, $\phi $ is azimuthal angle, $w_{0}$ is beam
half-width, and $t(\phi )$ is transmission function of the SLM at $z=0$. For
MFVBs, their initial phase structures with multiramp screw-edge dislocations
are assumed to be \cite{Gbur2016}
\begin{equation}
t(\phi )=\exp \left[ i\alpha (\phi -\frac{2\pi n}{m})\right] ,  \label{MTF}
\end{equation}%
where $m$ is an integer representing the number of multiramp mixed
screw-edge dislocations contained in their phase structures, and $%
n=\left\lfloor m\phi /(2\pi )\right\rfloor $ with $\left\lfloor \cdot
\right\rfloor $ the floor function. This function can be expend into the
following Fourier series%
\begin{equation}
t(\phi )=\sum\limits_{q=-\infty }^{\infty }c_{qm}\exp [iqm\phi ],
\label{Expension}
\end{equation}%
with $c_{qm}=\frac{\sin (\alpha \pi /m)}{\pi (\alpha /m-q)}\exp (i\alpha \pi
/m)$ and $q=\cdots ,-2,-1,0,1,2,\cdots $. Therefore, the output light fields
of such MFVBs in paraxial optical systems can be seen as a superposition of
a series of integer vortex beams, as follows%
\begin{equation}
E_{\text{out}}(\rho ,\varphi ,z)=\sum\limits_{q=-\infty }^{\infty }c_{qm}E_{%
\text{o,}qm}(\rho ,\varphi ,z),  \label{Eout}
\end{equation}%
where $\left( \rho ,\varphi ,z\right) $ are the coordinates at output plane,
and $E_{\text{o,}qm}(\rho ,\varphi ,z)$ is the output light field for an
integer Gaussian vortex beams with charge number $qm$. For a general optical
systems \cite{Collins1970,Zhao2000}, $E_{\text{o,}qm}(\rho ,\varphi ,z)$ can
be calculated as in Ref. \cite{WJS2019} and be given by%
\begin{widetext}
\begin{align}
E_{\text{o,}qm}(\rho ,\varphi ,z)& \!=\!\frac{(-i)^{|qm|\!+\!1}z_{R}^{2}}{%
(B\!-\!iAz_{R})^{3/2}}(\frac{\pi \rho ^{2}}{4Bw_{0}^{2}})^{1/2}\exp \left[
i(kL\!+qm\varphi )\right]   \notag \\
& \!\times \!\exp \left[ \frac{ik\rho ^{2}}{2B}\left( D\!+\!\frac{iz_{R}}{%
2(B\!-\!iAz_{R})}\right) \right]   \notag \\
& \!\times \!\left[ I_{\frac{|qm|\!-\!1}{2}}\left( \frac{z_{R}^{2}\rho
^{2}/w_{0}^{2}}{2B(B\!-\!iAz_{R})}\right) \!-\!I_{\frac{|qm|\!+\!1}{2}%
}\left( \frac{z_{R}^{2}\rho ^{2}/w_{0}^{2}}{2B(B\!-\!iAz_{R})}\right) \right]
,  \label{En_out}
\end{align}%
\end{widetext}
where $z_{R}=kw_{0}^{2}/2$ is the Rayleigh distance of initial beam, $L$ is
the eikonal along the propagation axis, $k$ is the wave number, and $A$, $B$%
, and $D$ are the elements of a $2\times 2$ ray transfer matrix $\left(
\begin{smallmatrix}
A & B \\
C & D%
\end{smallmatrix}%
\right) $ describing a linear optical system. When $m=1$, Eq. (\ref{Eout})
reduces to the case of FVBs with finite beams discussed in Ref. \cite%
{WJS2019}, while when $m=2,3,\cdots $, multiple screw-edge dislocations
leads to complex evolution of such MFVBs. By using Eq. (\ref{Eout}), we can
analytically obtain their light field evolution.

\begin{figure*}[hbtp]
\centering
\includegraphics[width=9cm]{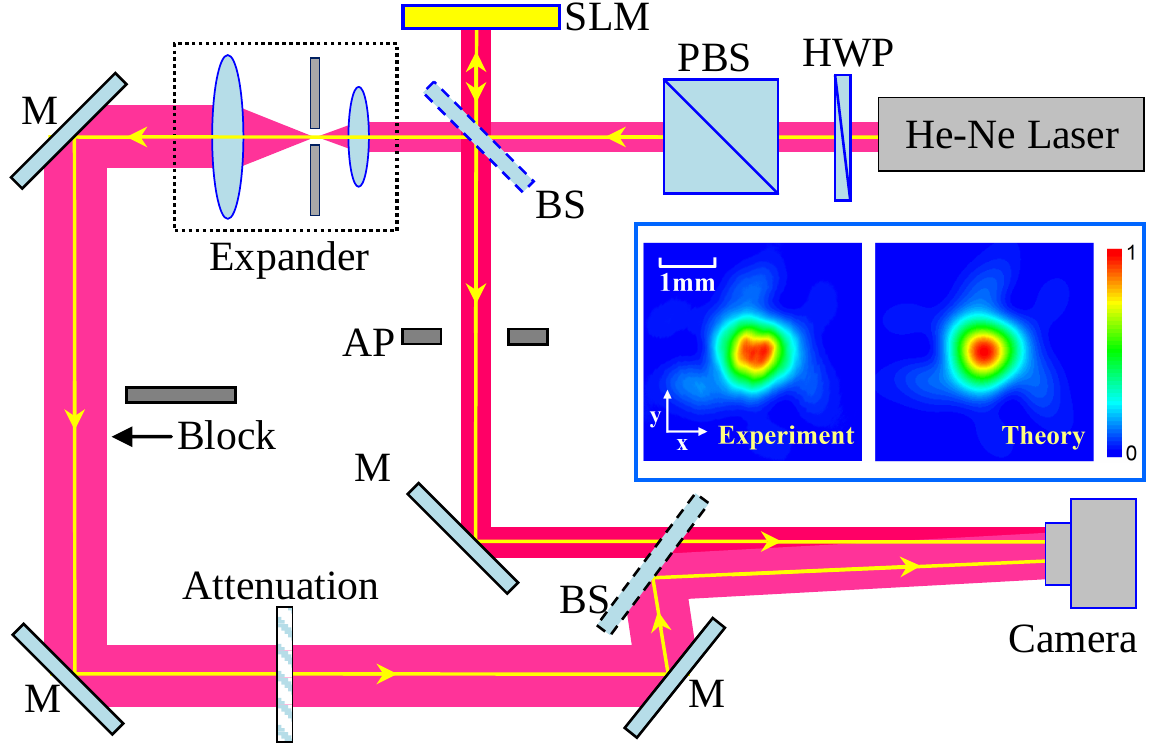}
\caption{Experimental setup for measuring intensity distributions of MFVBs
and their interference patterns with a wide Gaussian beam. Inset shows the
measured intensity of the MFVBs with $m=3, \protect\alpha=1$ and the
corresponding theoretical result (at $z=z_{R}$). Other notations are: HWP,
half-wave plate; BS, beam splitter; M, mirror; PBS, polarized beam splitter;
SLM, spatial light modulator; AP, Aperture. }
\label{Fig1}
\end{figure*}
Furthermore, in order to describe the whole property of such MFVBs, it is
very important to know the change of their total vortex strength, which is
defined by%
\begin{widetext}
\begin{align}
S_{m,\alpha }& =\frac{1}{2\pi }\underset{\rho \rightarrow \infty }{\lim }%
\int_{0}^{2\pi }d\varphi \partial \lbrack \arg E_{\text{out}}(\rho ,\varphi
,z)]/\partial \varphi  \notag \\
& =\frac{1}{2\pi }\underset{\rho \rightarrow \infty }{\lim }\int_{0}^{2\pi
}d\varphi \text{Re}[(-i)E_{\text{out}}^{-1}(\rho ,\varphi ,z)\partial E_{\text{%
out}}(\rho ,\varphi ,z)/\partial \varphi ].  \label{VS}
\end{align}%
\end{widetext}
In calculation, the value of $\rho $ in the above integral may influence the
result so that it should be sufficient large. By calculating Eq. (\ref{VS}),
one can obtain the change of the total vortex strength of such MFVBs in free
space, for which its transfer matrix is simply given by $\left(
\begin{smallmatrix}
1 & z \\
0 & 1%
\end{smallmatrix}%
\right) $ with $z$ being the propagation distance.

\section{Experimental results and discussions}
\begin{figure*}[hbtp]
\centering
\includegraphics[width=13cm]{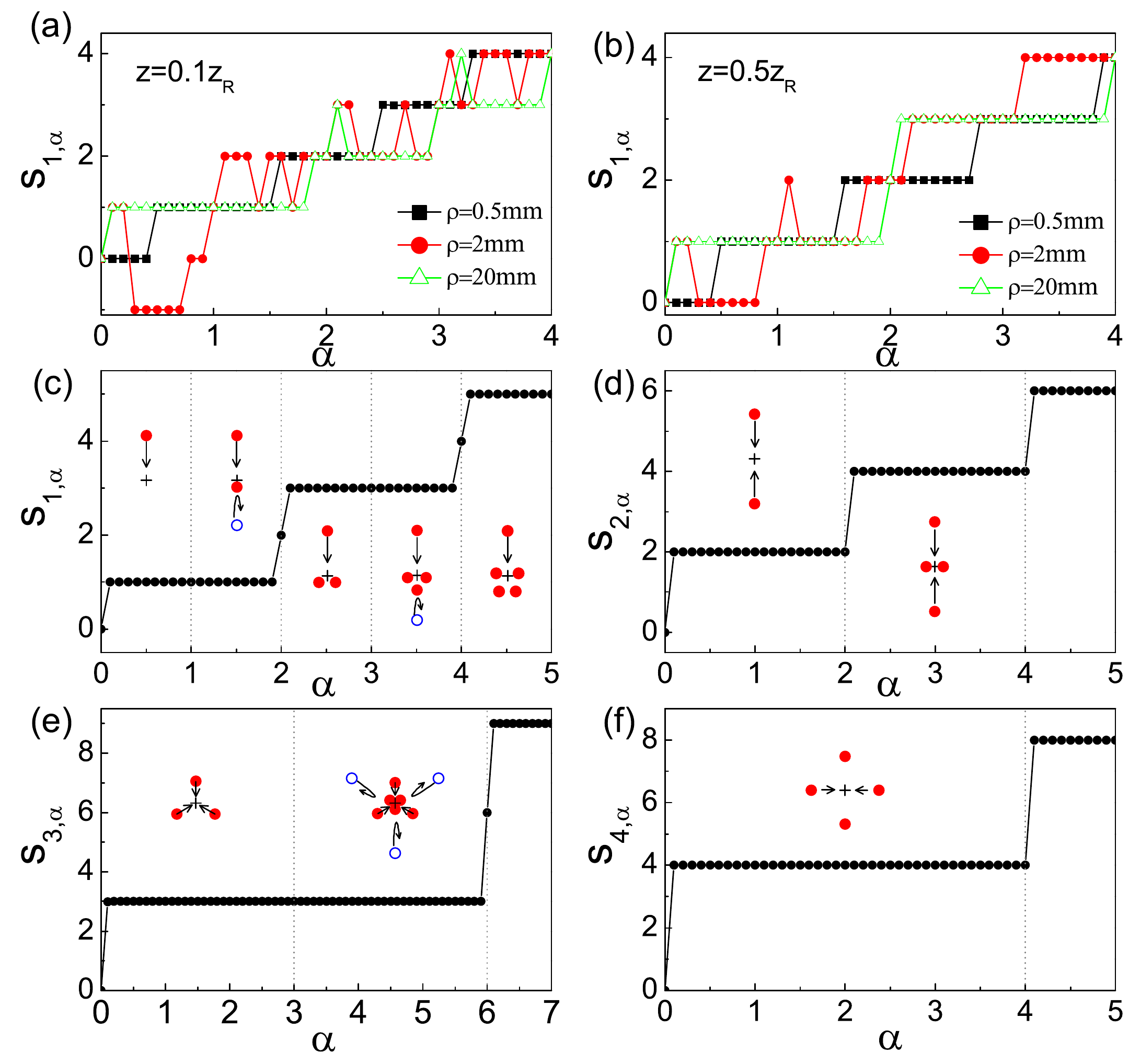}
\caption{Total vortex strength of MFVBs as a function of topological charge $%
\protect\alpha $ under different $m$. (a-b) The value of $S_{1,\protect%
\alpha }$ as a function of $\protect\alpha $ at (a) $z=0.1z_{R}$ and (b) $%
z=0.5z_{R}$ with different radius of the integral loop $\protect\rho =0.5$
mm, $2$ mm and 20 mm. (c-f) The value of $S_{m,\protect\alpha }$ as a
function of $\protect\alpha $ at $z=z_{R}$ with (c) $m=1$, (d) $m=2$, (e) $%
m=3$, and (f) $m=4$ under $\rho\rightarrow\infty$. The insets in (c-f) show the layouts of vortex movement
and generation with the increase of $\protect\alpha $ at far-field region,
where the red-solid and blue-open circles, respectively, denote the vortices
with +1 and -1 charge. The arrows indicate the movement direction of new
vortices when $\protect\alpha $ increases, and the cross symbol is the
geometric center of light field.}
\label{Fig2}
\end{figure*}
Experimentally, as shown in Fig. 1, MFVBs are generated by a
phase-modulation SLM (Holoeye PLUTO-2-VIS-056). The linearly-polarized beam
from a He-Ne laser (with $\lambda =633$ nm) first passes through a half-wave
plate and a polarized beam splitter, which leads to the intensity of light
being well controlled. In our experiment, the measured beam half-width of
the initial horizontally-polarized Gaussian beam is $w_{0}\approx 0.57$ mm,
which is also used in our theoretical calculation. Through a beam splitter,
the incident Gaussian beam splits into two parts. One is reflected into the
SLM for generating the designed MFVBs, which is selected from the first
order of the modulated beam by using an aperture. The other part is expanded
by using two confocal lenses to become a wide-waist Gaussian beam, which is
used to interfere with MFVBs, with a crossed angle, for the purpose to
verify vortex structures of such MFVBs. When light in the second path is
blocked, one can investigate the evolution of such MFVBs at different
distance of free space. All intensities are recorded by a CCD camera with
high resolution and 14-bit data depth. The inset of Fig. 1 shows both the
experimental and theoretical intensity distributions for the MFVB with $m=3$
and $\alpha =1$ at the position of $z=z_{R}$. It is clearly seen that the
measured intensity profile is in good agreement with the theoretical result.
Thus we only plot the measured intensities below, and the more comparison between
the experimental and theoretical distributions can be found in Appendix A (Appendix A, Fig. A1).

\begin{figure*}[hbtp]
\centering
\includegraphics[width=17.5cm]{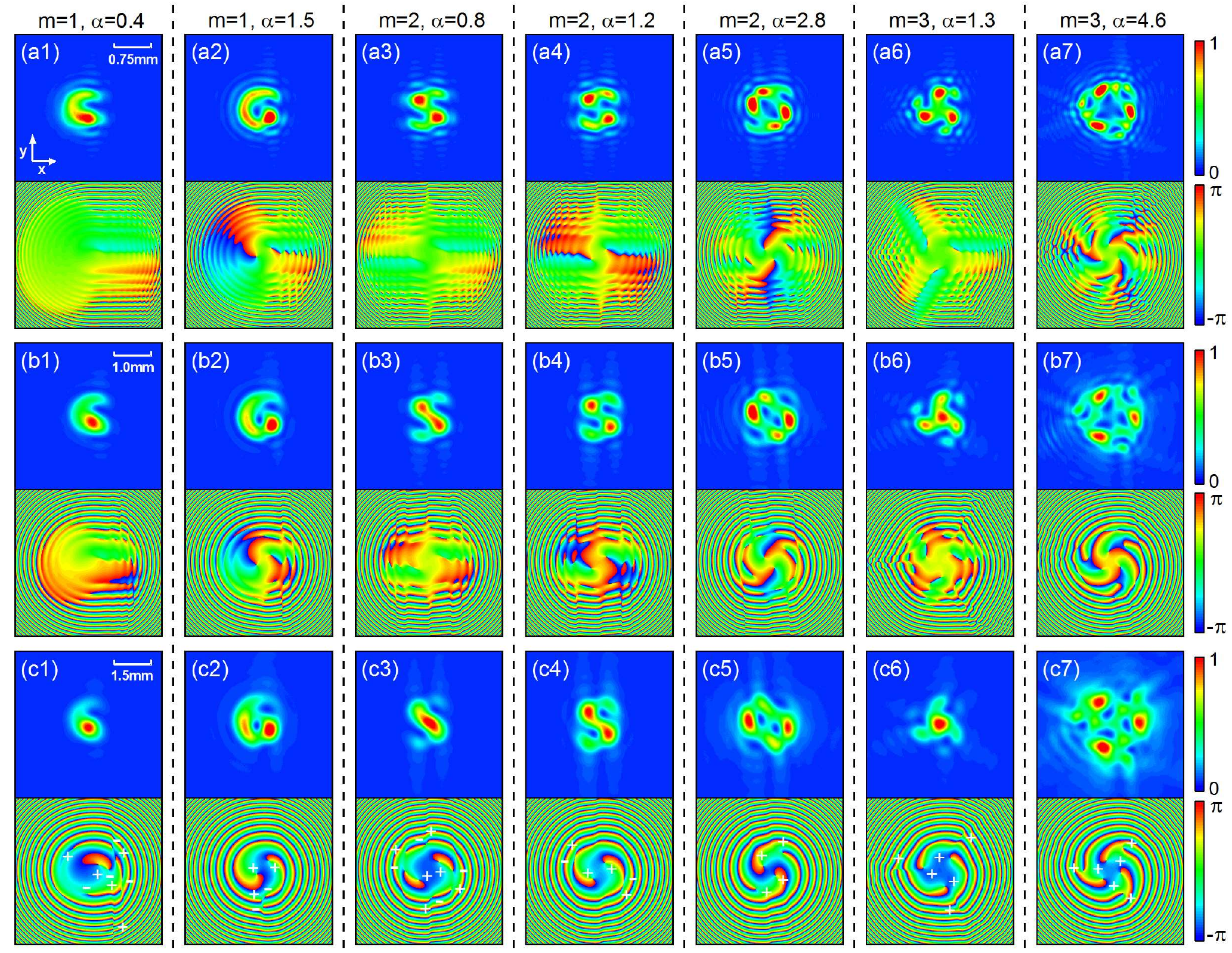}
\caption{Experimental measured intensities (upper) and
corresponding theoretical phase (down) of the MFVBs with different $m$ and $\protect%
\alpha $ at different positions (a1-a7) $z=0.1z_{R}$, (b1-b7) $z=0.2z_{R},$ and (c1-c7) $z=0.5z_{R}.$ Phase distributions in (c6, c7) only show the central six charge +1 vortices while other three charge -1 vortices located outer region are not shown (please see the Figs. A3(a) and (d) in Appendix C). In (c1-c7), the positive unit vortex is signed by ``+", the negative unit vortex is signed by ``-".}
\label{Fig3}
\end{figure*}
Figure \ref{Fig2} shows the dependence of total vortex strength $S_{m,\alpha
}$ on the parameter of topological charge $\alpha $ of MFVBs in free space.
As mentioned above, when $m=1$, it reduces to the cases of
conventional FVBs. In our recent work \cite{WJS2019}, we have clearly
demonstrated that total vortex strength for an FVB occurs a unit jump only
when $\alpha $ is before and after (but very close to) any even integer
number due to two different mechanisms for generation and movement of
vortices in focusing systems, which correspond to the cases of far-field
region (i.e., Fraunhofer diffraction region). Here we further show that, for
a practical FVB (i.e., the incident light fields with finite beam-width),
its $S_{m,\alpha }$ is, in principle, a constant during its propagation in
free space. However, due to the complex evolution and unstable phase
structures in the near-field region \cite%
{Basistiy2004,Leach2004,Fang2017} (roughly when the propagation
distance is smaller than $0.5z_{R}$), the value of $S_{m,\alpha }$ for
practical FVBs or MFVBs are dependent on the integral loop of Eq. (\ref{VS})
(i.e., the value of $\rho $), see Figs. 2(a) and 2(b). From Figs. 2(a) and
2(b), it is clearly seen that $S_{1,\alpha}$ are unstable because in the
outer regions of the beams there are a series of positive and negative
vortices (see Fig. 3), which may be included or not included when the loop
is small. When the loop is sufficient large and the propagation distance
increase, $S_{1,\alpha }$ trends to be stable and finally becomes a constant
during propagation, see the cases of $\rho =20$ mm at $z=0.5z_{R}$ in Fig.
2(b).

\begin{figure*}[hbtp]
\centering
\includegraphics[width=17.5cm]{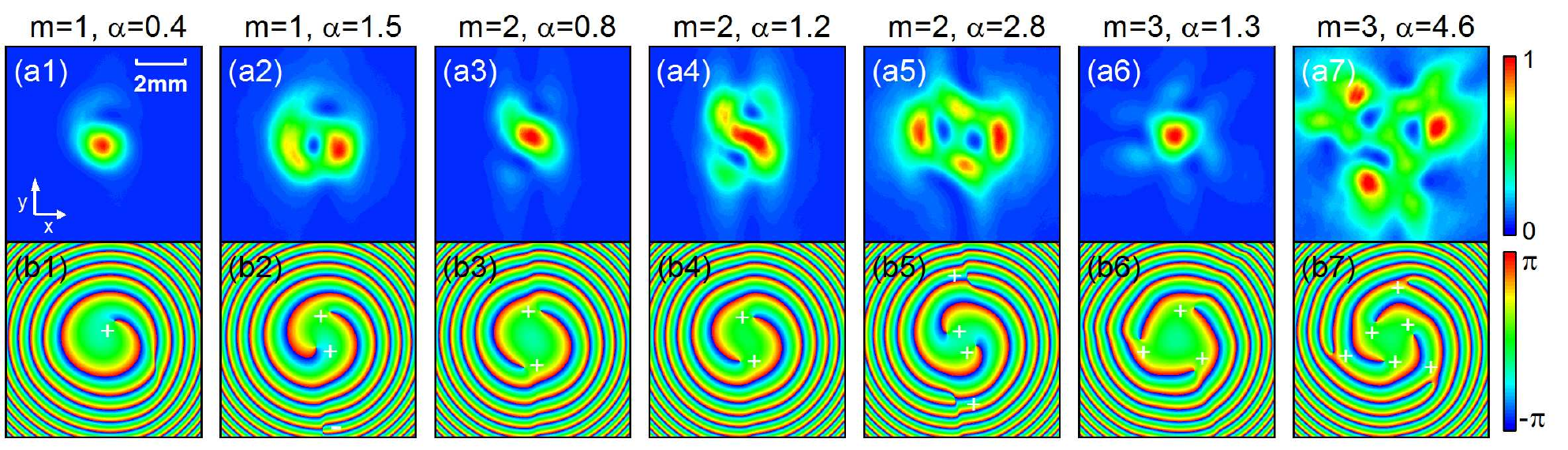}
\caption{Experimental measured intensities (a1-a7) and corresponding
theoretical phase structures (b1-b7) of the MFVBs with different $m$ and $\protect\alpha
$ at $z=1.5z_{R}$. Phase in (b7) also only shows the central six charge +1 vortices while there are also three charge -1 vortices located outer region (see the Fig. A3(f) in Appendix C). In (b1-b7), the positive unit vortex is signed by ``+", the negative unit vortex is signed by ``-".}
\label{Fig4}
\end{figure*}

In Figs. 2(c-f), it shows the steady change of $S_{m,\alpha }$ at $z=1.0z_{R}
$ under the integral with a sufficient large loop of $\rho \rightarrow
\infty $. Clearly, for practical FVBs or MFVBs in free space, when $m$ is
an even number, the value of $S_{m,\alpha }$ jumps with amplitude $m$
when $\alpha $ is slightly larger than any even integer $lm$ with $%
l=0,\pm1,\pm2\cdots $ also being an integer. When $m$ is an odd number, its value
of $S_{m,\alpha }$ first jumps at the place where $\alpha $ is slightly
before any $2lm$, and then $S_{m,\alpha }$ exactly equals to $2lm$ at $%
\alpha =2lm$, and there is another kind of jumps happening when $\alpha $ is
slightly larger than $2lm$. All jumps have the amplitude $m$.

Meanwhile, we would like to emphasize that Berry's results for FVBs in Ref.
\cite{Berry2004} can be recovered under the ideal case that the incident
light is an ideal plane wave. Or say, one can approximately mimic Berry's
results when the incident beam has a wide beam-width in the near-field
region, with an optimal integral loop. However, in principle, the total
vortex strength of any practical FVBs or MFVBs with finite beam-widths
always obeys the following relation

\begin{equation*}
S_{m,\alpha }=\left\{
\begin{array}{l}
(2l-1)m\text{,}\ \text{\ for }2m(l-1)<\alpha <2ml,\text{\ }m=1,3,5\cdots ,
\\
2ml,\ \ \ \ \ \ \ \text{for }\alpha =2ml,\text{\ }m=1,3,5\cdots , \\
ml,\ \ \ \ \ \ \ \ \text{for }m(l-1)<\alpha \leq lm,\text{\ }m=2,4,6\cdots .%
\end{array}%
\right.
\end{equation*}%
The above rule for MFVBs is also different from the result in Ref. \cite%
{Gbur2016}, which are only suitable for the ideal case that the incident
light is the plane wave.

In order to understand the above rule of total vortex strength of practical
FVBs and MFVBs, it is important to observe the intensity and phase evolution
of these beams in both the near-field and far-field regions of free space.
In Fig. \ref{Fig3}, it shows clearly that there are open dark structures in
the near-field regions of FVBs with $m=1$ when $\alpha $ approaches to half
integer (or $m/2$ when $m=2,3,\cdots$). These open dark structures are actually corresponding to the
vortex-pair generation or pairs of positive and negative vortices \cite%
{Berry2004,Leach2004}, see the phase distributions in Figs. 3(a1) and 3(b1). These positive and/or negative vortices have complex evolution in
near-field regions. However, as the propagating distance $z$ increases,
through a careful observation, almost all vortices first generate, then evolute, and finally
annihilate with their nearest opposite-sign vortices (see Appendix B, Fig. A2). This property is actually the same as the evolution of those
optical vortex knots \cite{Dennis2010}. In the far-field region, all
remained vortices are stable and may rotate an angle of $\pi /2$ due to Gouy
phase \cite{Vasnetsov1998}. Meanwhile, as $z$ is close to the far-field
region, there exist only a finite number of vortices, from which one can
easily determine the value of $S_{m,\alpha }$.

For examples, for MFVBs with $m=2$ and $m=3$, in Fig. 3, it also shows how the beams
evolute from the near-field to far-field regions. There are more complex
intensity and phase structures, and pairs of vortices form, evolute, and
annihilate in near-field regions. This is the reason that a finite integral
loop used in Eq. (\ref{VS}) is hard to find a correct result in near-field
regions. However, as the same as the cases of $m=1$, all remained vortices
for MFVBs also become stable as $z$ approaches the far-field regions.

\begin{figure*}[htbp]
\centering
\includegraphics[width=14cm]{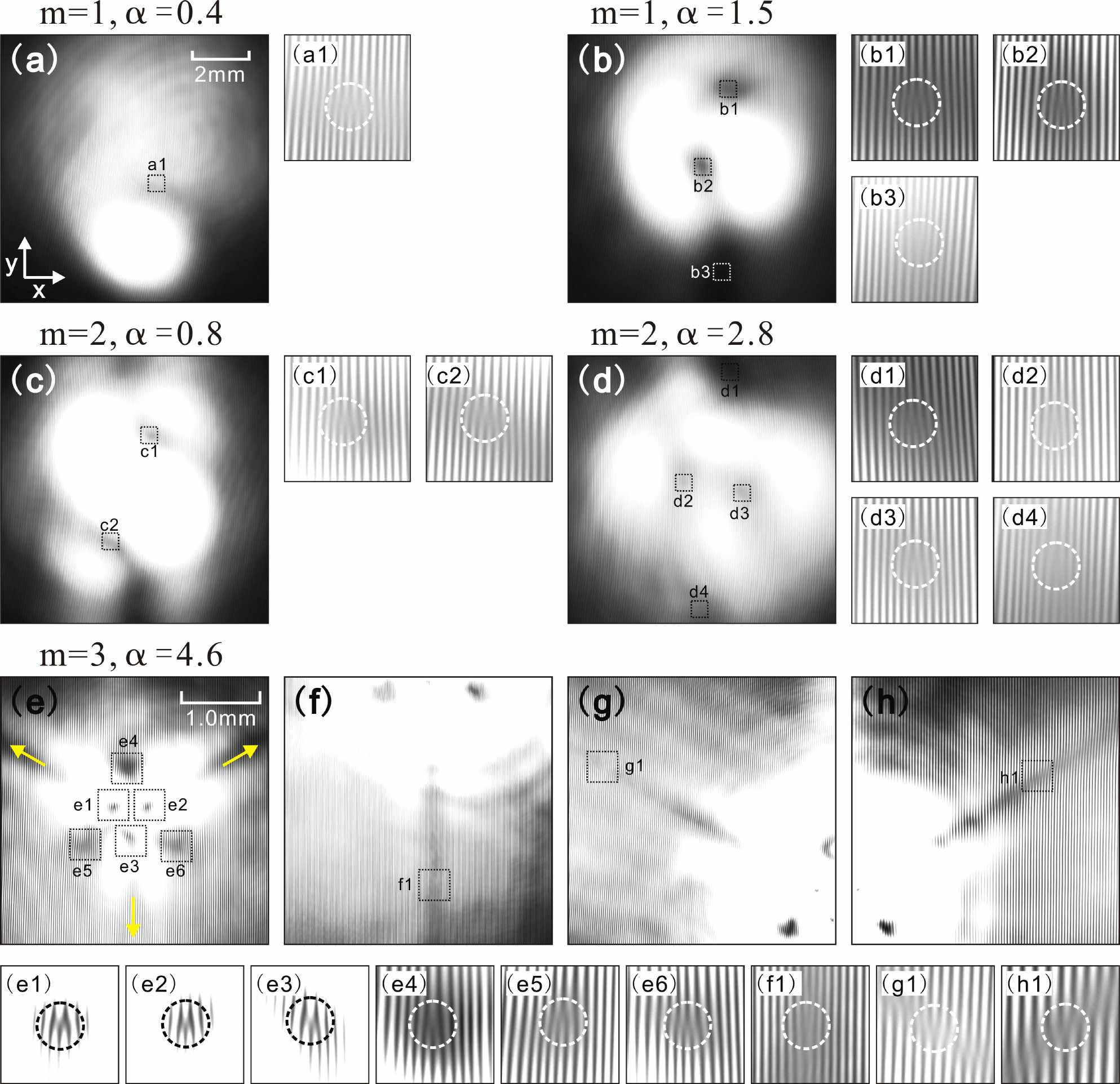}
\caption{Measured interference patterns of MFVBs with a titled wide-width Gaussian beam.
 The different interference patterns for (a) $m=1, \alpha=0.4$, (b) $m=1, \alpha=1.5$, (c) $m=2, \alpha=0.8$, and (d) $m=2, \alpha=2.8$  are measured at $z=1.5z_{R}$.
  (e)-(h) The interference patterns are measured at the focal plane of a 2-$f$ lens system with a focal length 50 cm for the case of $m=3, \alpha=4.6$, and here (f, g, h) are the outer regions of (e) in three directions (indicated by yellow  arrows).
 All fork-like patterns marked by square boxes in (a-h) are enlarged and marked by dash circles in the corresponding subfigures.}
\label{Fig5}
\end{figure*}
Figure 4 further demonstrates the measured intensity distributions and their
corresponding phase structures for FVBs and MFVBs at the far-field region ($%
1.5z_{R}$). By including the vortex sign, the value of total vortex strength
is coincidence with the prediction of Figs. 2(c-f), and the distributions of
stable vortices for practical MFVBs at the far-field region are also
displaced in Fig. 4. For example, when $m=2$ and $\alpha =0.8$, it is
observed that there already exist two +1 vortices. But when $m=2$ and $%
\alpha =1.2$, there are still two +1 vortices to be observed. This result is
different from that in Ref. \cite{Gbur2016}, where the value of $S_{m,\alpha
}$ jumps when $\alpha =m/2$. There are similar situations for MFVBs with $m=3
$ (also see Appendix C). It should be emphasized that the
vortices in outer low-intensity area are not only important to $S_{m,\alpha }
$ but also to the issues about higher transverse-vector components of beams.

For practical FVBs with $m=1$, the rule of vortex movement and generation as
a function of $\alpha $ is schematic in Fig. 2(c). In Figs. 2(d)-2(f), it
further shows the layout of vortex movement and generation with the increase
of $\alpha $. All these results are summarized from phase distributions
around a circle loop with a very large radius at far-field regions (see the
Appendix C). These rules are important to understand the
evolution of vortices for real MFVBs and FVBs in far-field regions.

In order to obtain direct evidences on total vortex strength and phase
structures of MFVBs, we have performed the interference experiments by using
these typical MFVBs with a titled wide-width Gaussian beam, shown in Fig. 1,
when the optical path after the first beam splitter is not blocked. From the
above discussions, we know that the light fields (containing vortices) at
far-field zone are stable. Therefore, the interference phenomena are
recorded by the CCD camera at those far-field regions. Figure 5 shows some
typical interference patterns at the plane of $1.5z_{R}$, from which one can
identify all vortices indicated by fork-like interference patterns. The
opening direction of fork-like pattern determines the sign of vortex. As
shown in Fig. 5(a), for an FVB with $m=1$ and $\alpha =0.4$, there is only
one downward fork-like pattern detected, and its feature shows a vortex with
+1 charge existing in the field. When $m=1$ and $\alpha =1.5$, one can find
three fork-like interference patterns on the image, however two of them are
downward and the other is upward. This tells us that there are two vortices
with +1 charge and one vortex with -1 charge, thus one still have $%
S_{1,1.5}=1$ in this case. For MFVBs with $m=2$ and $m=3$ in Figs. 5(c, d,
e), respectively, through a carefully experimental observation (also see
Appendix D), we can conclude that both the number and sign of
all vortices for MFVBs in far-field regions are verified by using such
interference patterns. Here we emphasize that when $m$ is an odd number, there are possible opposite-sign vortices existed in the outer extremely-low intensity region at far-field zone, which makes the interference measurement become very difficult, however, by using 2-$f$ lens system, it will reduce the difficulty to measure these vortices at focal plane. The measured result of total vortex strength on MFVBs
and FVBs is in good agreement with the theoretical prediction of Fig. 2.

\section{Summary}
In a conclusion, we have studied the total vortex strength and vortex
structures of MFVBs in practical situations which are generated by using a
finite Gaussian beam not an ideal plane-wave light field. In principle, when
MFVBs propagate in free space, their total vortex strength is always
conserved. However, due to the unstable properties of intensity and phase
structures in near-field region, it is better to determine the value of
total vortex strength versus the topological charge for such MFVBs in
far-field zone. Our result offers a rule of the total vortex strength as a
function of topological charge for such practical MFVBs, which is different
from the results of both the ideal plane-wave FVBs \cite{Berry2004} and
MFVBs \cite{Gbur2016}. Although one might mimic the ideal results in
near-field zone by using the wide-width light fields, the essential
propagating properties of practical MFVBs and FVBs in free space are
disclosed to have totally different behaviors, especially in far-field
regions, which are never observed before. Our result also suggests that it
is more convenient to directly evaluate the total vortex strength at
far-field zone, where all vortices are stable and countable. Our result is
important for micro-particle trapping and provides an understanding for
complex fractional vortex structured light beams in practical optical
applications.

\begin{acknowledgments}
This work was supported by the Fundamental Research Funds for the Center Universities (No. 2017FZA3005); National Key Research and Development Program of China (No. 2017YFA0304202); National Natural Science Foundation of China (NSFC) (grants No. 11674284);
and the Fundamental Research Funds for the Center Universities (No.
2019FZA3005).
\end{acknowledgments}

\begin{appendix}
\setcounter{figure}{0}
\renewcommand{\thefigure}{A\arabic{figure}}
\section{Comparison of the measured intensity with the theoretical results for MFVBs propagating from near-field  to far-field regions}

\begin{figure*}[htbp]
\centering
\includegraphics[width=17cm]{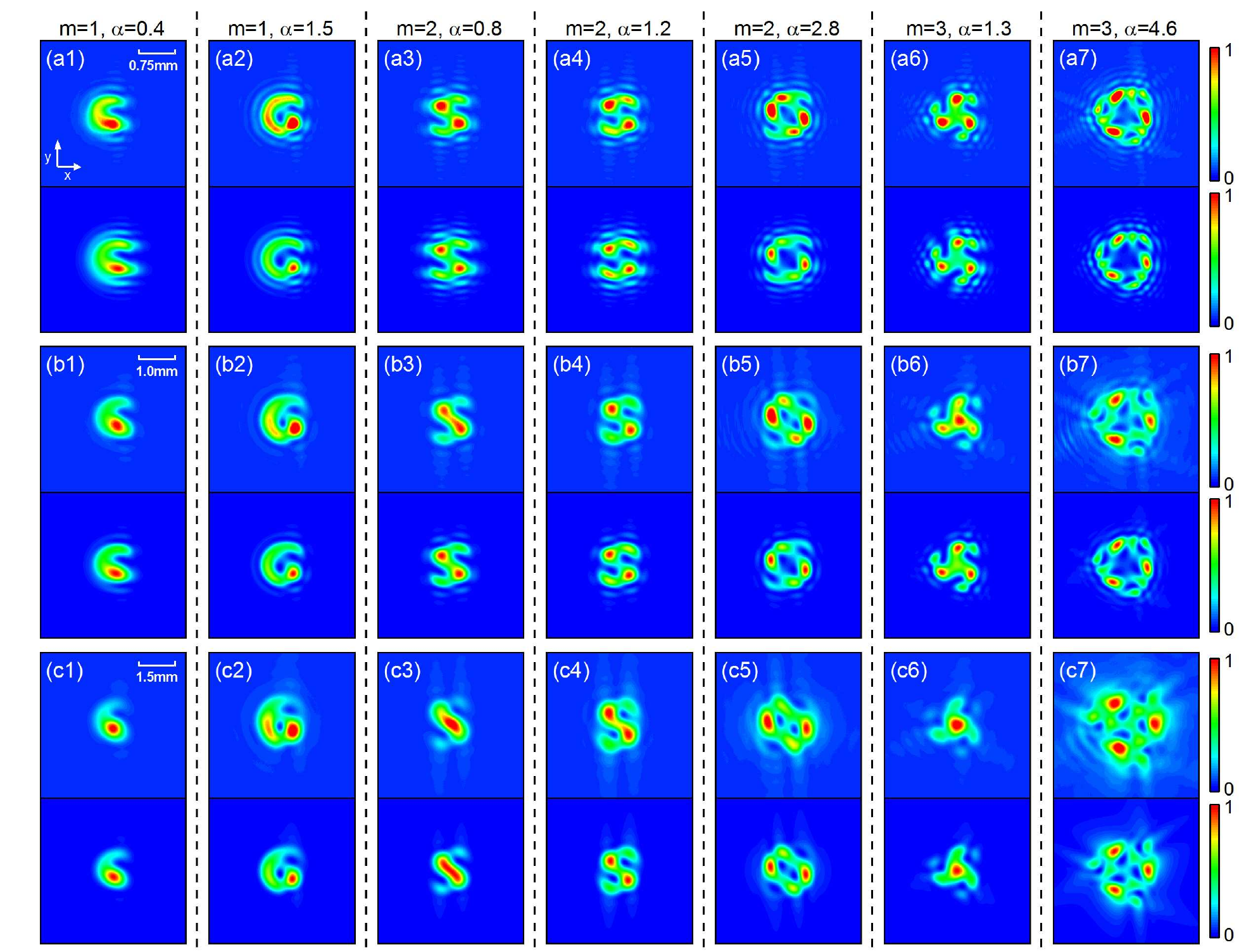}
\caption{Experimental measured (upper) and corresponding theoretical (bottom)
intensities of the MFVBs with different $m$ and $\alpha$ at different positions (a1-a7) $z=0.1z_{R}$, (b1-b7) $z=0.2z_{R},$ and (c1-c7) $z=0.5z_{R}.$}
\label{FigA1}
\end{figure*}
Figure A1 shows the experimental measured intensities and corresponding theoretical intensities of the MFVBs. The measured intensities are in good agreement with theoretical intensities for the MFVBs with different $m$ and $\alpha$. Therefore, the theoretical equation of Eq. (4) can describe the field evolution of such MFVBs propagating in free space and the phase vortex structure from Eq. (4) can explain the complex dark-field structures at different propagating distance.

\section{The evolution of vortices at near-field region}

\begin{figure*}[htbp]
\centering
\includegraphics[width=8cm]{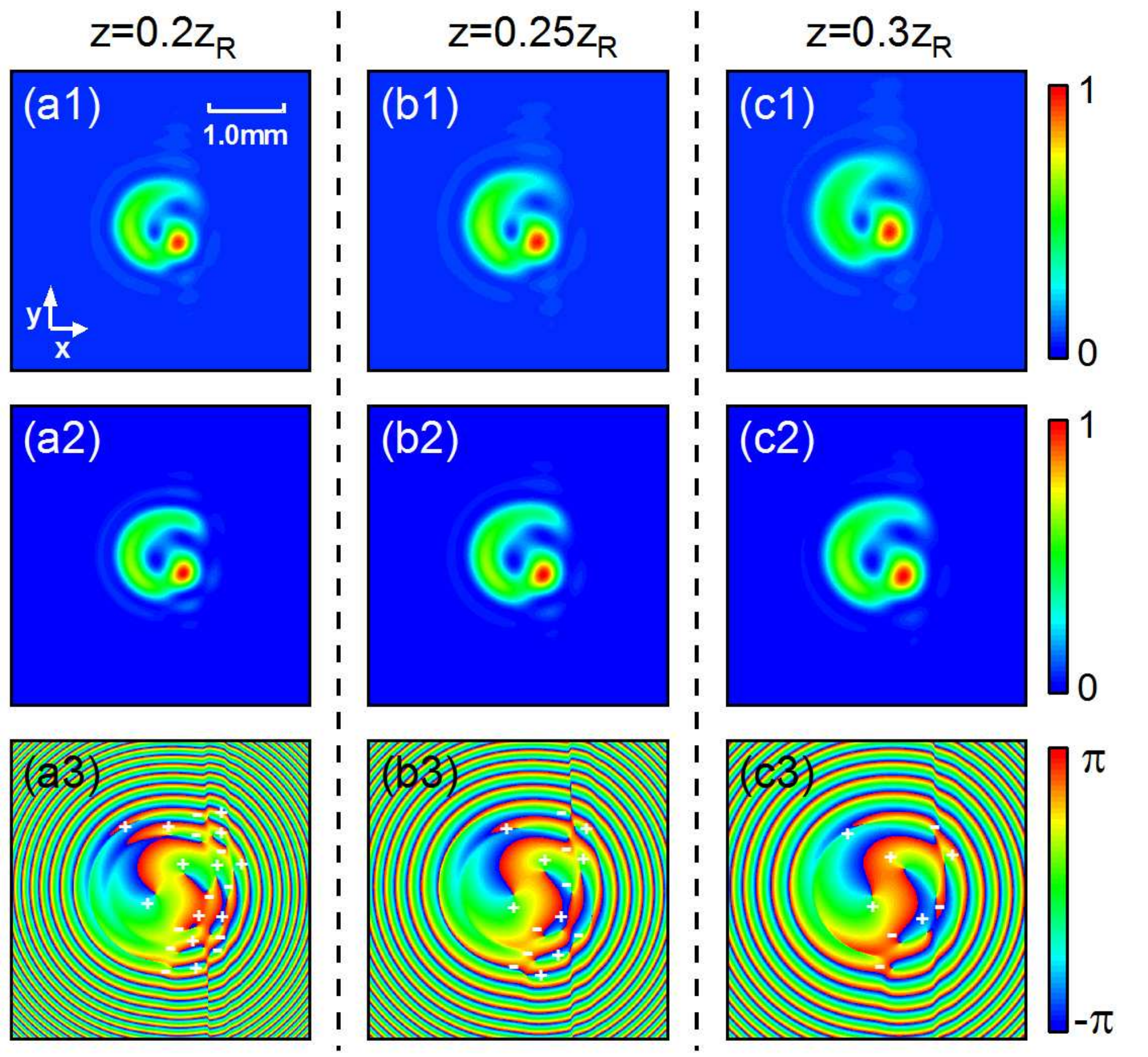}
\caption{Experimental measured intensities (a1-c1) and corresponding theoretical
intensities (a2-c2) and phase (a3-c3) of FVB with $m=1, \alpha=1.5$ at different positions (a) $z=0.2z_{R}$, (b) $z=0.25z_{R},$ and (c) $z=0.3z_{R}.$ In (a3-c3), the positive unit vortex is signed by ``+", the negative unit vortex is signed by ``-".}
\label{FigA2}
\end{figure*}
Figure A2 shows the evolution of the vortices for the FVB with $m=1, \alpha=1.5$ at near-field region. It is observed that the measured intensity is in a good agreement with the theoretical result. Furthermore, it is observed in Fig. A2(a3) that there are pairs of vortices with opposite charges at $z=0.2z_{R}$. As the beam propagates from $z=0.2z_{R}$ to $z=0.3z_{R}$, see Figs. A2(b3) and (c3), pairs of vortices with opposite charge annihilate each other.

\section{The phase of the MFVBs under a large $\rho$ at far-field region}

\begin{figure*}[hbt]
\centering
\includegraphics[width=16cm]{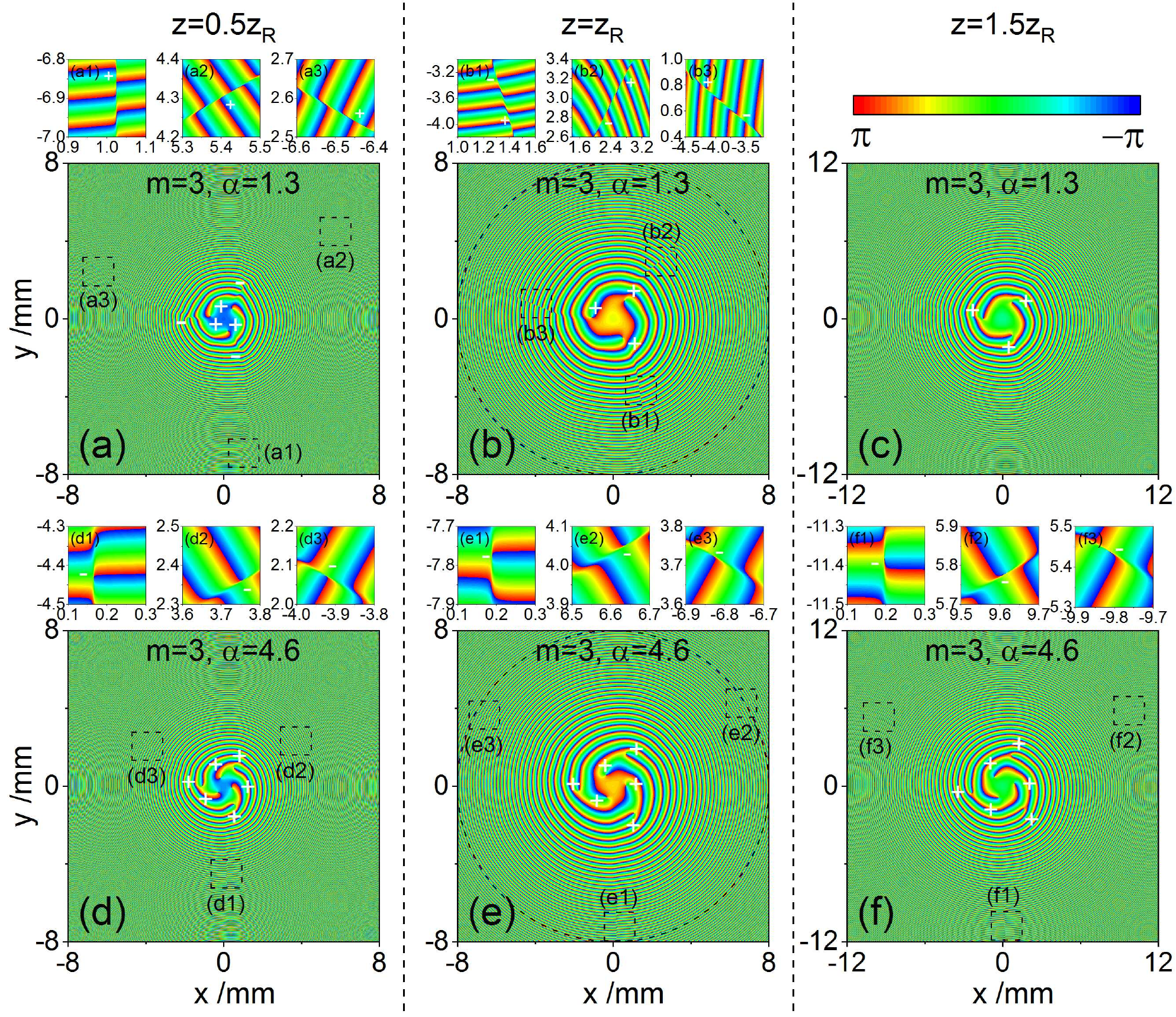}
\caption{The theoretical phase distributions of the MFVBs with $m=3, \alpha=1.3$ and $m=3, \alpha=4.6$ at different positions (a, d) $z=0.5z_{R},$ (b, e) $z=z_{R},$ and (c, f) $z=1.5z_{R}$ in far-field region. The vortices at the outer region marked by square boxes are
enlarged in the corresponding subfigures. In (a-f), the positive unit vortex is signed by ``+", the negative unit vortex is signed by ``-". The dash circles in (b, e) denote the loops with radius $\rho=8$ mm at phase plane.}
\label{FigA3}
\end{figure*}

\begin{figure*}[hbtp]
\centering
\includegraphics[width=12cm]{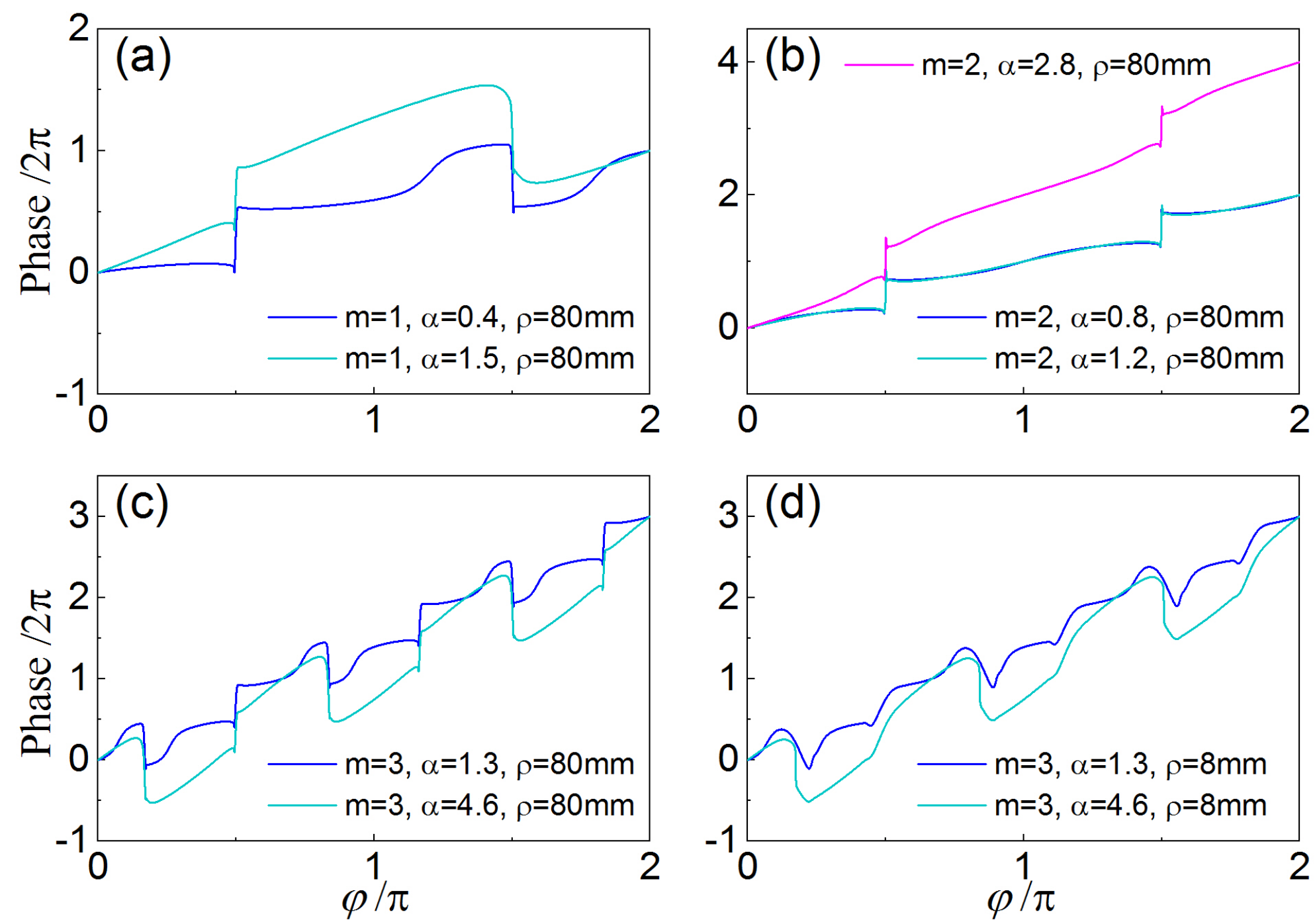}
\caption{The theoretical phase curves as a function of azimuthal angle $\varphi$ at the loop on phase plane with the radius (a, b, c) $\rho=80$ mm, (d) $\rho=8$ mm for MFVBs with different $m$ and $\alpha$. Here, the propagating distance $z=z_{R}$. The phase curves in (d) are corresponding to the phase marked by dash circles in Figs. A3(b) and (e).}
\label{FigS4}
\end{figure*}

\begin{figure*}[hbtp]
\centering
\includegraphics[width=13cm]{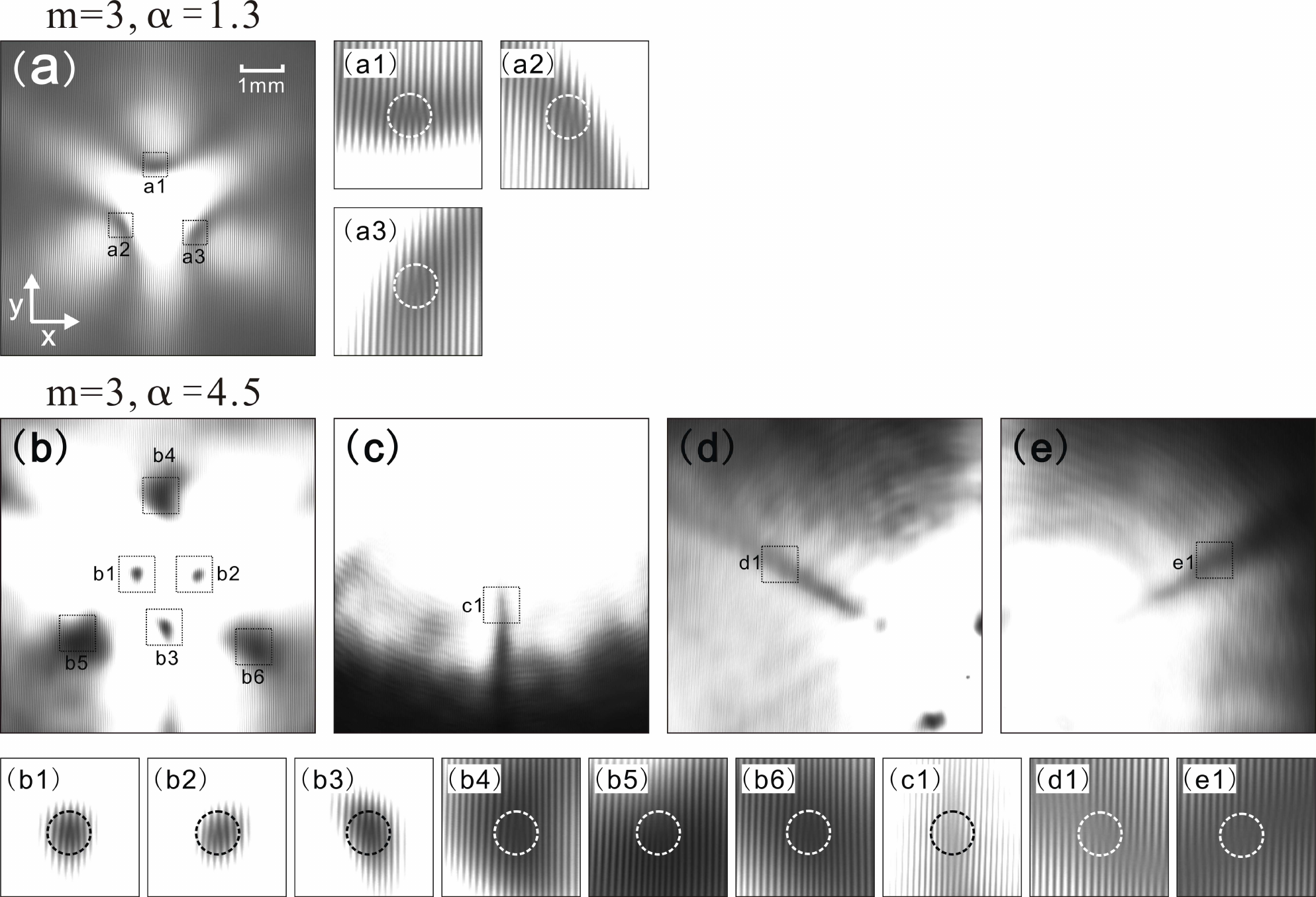}
\caption{Measured interference patterns of MFVBs with a titled wide-width Gaussian beam measured at focal plane of a 2-$f$ lens system with a focal length 50 cm for the case of (a) $m=3, \alpha=1.3$ and (b) $m=3, \alpha=4.5$. All fork-like patterns marked by square boxes in (a, b, c) are
enlarged and marked by dash circles in the corresponding subfigures.}
\label{FigA5}
\end{figure*}
Figure A3 shows the theoretical phase of some typical MFVBs within the region contained all vortices at different propagating positions. It is observed that in Fig. A3(a) that for MFVBs with $m=3, \alpha=1.3$ at $z=0.5z_{R}$ there are three charge +1 vortices at central region and three pairs of opposite-sign vortices with $\pm1$ charge at outer region. Therefore, the total vortex strength $S_{3,1.3}=3.$ As distance $z$ increases to 1.5$z_{R}$, see Figs. A3(b) and (c), the outer three pairs of unstable vortices annihilate each other. Finally the phase distribution becomes stable and the three charge +1 vortices remain. However, it is seen in Fig. A3(d) that for MFVBs with $m=3, \alpha=4.6$ at $z=0.5z_{R}$ there are nine vortices which is similar to the case of $m=3, \alpha=1.3$. These nine vortices maintain as the beam propagates to $z=1.5z_{R}$. These two different evolution behaviors of vortices between $m=3, \alpha=1.3$ and $m=3, \alpha=4.6$ due to two different mechanisms which have summarized in the insets of Fig. 2(e). Figs. A4(a, b, c) demonstrate the phase curves as a function of azimuthal angle $\varphi$ at a sufficient large loop on phase plane with the radius $\rho=80$ mm at $z=z_{R}$. These results verify the total vortex strength in Fig. 2. In addition, Fig. A4(d) shows the phase as a function of azimuthal angle $\varphi$ under $\rho=8$ mm at $z=z_{R}$. As compared with the cases of phase at $\rho=80$ mm, the accumulation of the phase is all equal to 6$\pi$ on both the loop of $\rho=8$ mm and $\rho=80$ mm. It tells us that the phase in Figs. A3(b) and (e) have already contained all the vortices and total vortex strength is equal to 3.

\section{The measured interference patterns at far-field region }
Figure A5 give more experimental results of the interference patterns of MFVBs with a title wide-width Gaussian beam measured at far-field region. It is seen in Fig. A5(a) that there are three downward fork-like pattern detected, which indicates that there are three +1 vortices for $m=3, \alpha=1.3$. For $m=3, \alpha=4.5$, there are total nine fork-like patterns, six of which are downward while the other three are upward. Therefore, the total vortex strength $S_{3,4.5}=3$. It confirms again our prediction of total vortex strength in Fig. 2.
\end{appendix}

\end{document}